
\input phyzzx
\hsize=6.0in
\vsize=8.9in
\hoffset=0.0in
\voffset=0.0in
\overfullrule=0pt

\FRONTPAGE
\line{\hfill BROWN-HET-894}
\line{\hfill TA-492}
\line{\hfill February 1993}
\vskip1.0truein
\titlestyle{{FIELDS AND SYMMETRIES OF 2D STRINGS}
 \foot{Work supported in
part by the Department of Energy under
contract DE-FG02-91ER40688-Task A}}
\bigskip
\author{Antal JEVICKI}
\centerline{{\it Department of Physics}}
\centerline{{\it Brown University, Providence, RI 02912, USA}}
\bigskip
\centerline{(An invited talk at the Nishinomiya--Yukawa Memorial}
\centerline{Symposium, Japan, November 1993)}
\abstract

\endpage

\chapter{Introduction}

Studies of strings in lower dimensions have produced extensive results
with major new insights.  First of all, the fruitful
relationship to the 1 dimensional matrix model [1]
has provided exact solutions for correlation functions and free
energy.  The dynamics of a sole field theoretic degree of freedom,
the massless tachyon has been well understood.  It is given by a
simple scalar Lagrangian of the collective field [2] theory.  In this
field theory one can expand perturbatively in the string coupling
constant and study loop effects [3].  This field
theory is also exactly integrable possessing an infinite sequence of
conserved currents and an exact S--matrix [4-8].  In addition to the
tachyon there also appears an infinite sequence of imaginary energy
discrete states originating in the inverted matrix oscillator.
The S--matrix can also be
found using the simple dynamics of the harmonic oscillator as we
explain.  It is in the matrix model language that a larger
space--time symmetry appears naturally and was seen [8] to take the
form of a $W_{\infty}$ algebra.  It plays a role of a spectrum generating
algebra leading to the infinite sequence of discrete states.
The exact solution of the theory can be traced to this symmetry.
The same symmetry was also established in the
conformal field theory approach [11].  Vertex operators for discrete states
close under operator products with structure constants of the
$W_{\infty}$ group.  The symmetry charges can be constructed. They
act nonlinearly on the tachyon (implying nontrivial Ward identities).
This nonlinear identities (which are sufficient to deduce the
S--matrix) can be seen to be the nonlinear collective
field representations [13].  It is in this sense that one can
think of the scalar field theory as a (minimal) realization of the
$W_{\infty}$ structure.  An extension of this would include a sequence
of additional discrete fields (representing the global degrees of
freedom).  The problem of understanding these fields, their couplings
to the tachyon and the complete gauge invariant theory is nontrivial.
We address some aspects of this problem.  We first describe
in Section 2 the simple correspondence
of the scalar field theory and the matrix model.  This we explain
can be used to deduce exact eigenstates and even the S-matrix directly from the
matrix hasmonic oscillator.  In section 3 we discuss $W_{\infty}$.
In section 4 we describe a possible
extension involving coupling of discrete topological fields to the
scalar tachyon.
\bigskip

\chapter{Scalar Dynamics}
\smallskip

The lowest excitation of string theory, the tachyon becomes massless in
two dimensions and represents the only real field theoretic degree of
freedom.  It is described by the scalar field $\phi (x,t)$ and its
conjugate $\pi (x,t)$ with its dynamics completely given by the cubic
(collective) field theory
$$H = \int {dx\over2\pi} \left\{ {1\over 6} \left(\alpha_+ (x,t)^3 -
\alpha_- (x,t)^3\right) - {1\over 2} x^2 \left( \alpha_+ - \alpha_-
\right) \right\}$$
Here $\alpha_{\pm} = \Pi_{,x} \pm \pi \phi$ simply denotes the two chiral
components.  The ground state is given by the static background $\pi
\phi_0 = \sqrt{x^2 -\mu}$ and the corresponding perturbation expansion
in $g_{st}^2 = 1/\mu^2$ defines the scattering amplitudes [3].
These were computed [3-6] explicitly and compared with similar amplitudes
in conformal field theory [7].
The agreement of S--matrix elements (and loop corrections)
implies the exactness of the above Lagrangian.

The fact that the collective Lagrangian is induced from the
simple matrix dynamics of an inverted oscillator is responsible for
the exact solvability [4,8] of the theory.  A set of simple rules gives the
transition:  the matrix variables are $M(t)$ and $P(t) = \dot{M}$ and
the transition to field theory can be summarized by:
$$\eqalign{ M & \rightarrow x\cr
P & \rightarrow \alpha (x,t)\cr
\Tr & \rightarrow \int {dx\over 2\pi} \int_{\alpha_{-}}^{\alpha_{+}}
d\alpha }$$
The matrix model hamiltonian
$$H = {1\over 2} \Tr \left( P^2 - M^2\right)$$
is then seen to transform into the cubic collective hamiltonian.  In
addition all commutation relations (or Poisson brackets) of U(N)
invariant observables simply go over
from the matrix model to the field theory.  More importantly a
remarkable physical phenomenon taken place:  from linear matrix model
dynamics
$$M(t) = M(0) ch\,t + P (0) sh\,t$$
a highly nonlinear string theory dynamics is generated as a
collective effect.

One can use the above transition
rules to directly compute
general scattering amplitudes and exhibit higher symmetries.  First a
one parameter set of exact states [8] is generated from the oscillators
$$\Tr \left( P\pm M\right)^n\rightarrow
 T_n^{\pm}  = \int {dx\over 2\pi} \, {(\alpha\pm x)\over
n+1}^{n+1} $$
$$ \left\{ H, T_n^{\pm} \right\} = \pm n T_n^{\pm}$$
They correspond to exact tachyon eigenstates $(p_o = E = \pm in)$
and represent field theoretic versions of tachyon vertex operator
of the conformal field
theory
$$T_p^{\pm} = e^{ipX^{0} + (-2 + \vert p \vert )\varphi}$$
We now give a simple derivation of the general tachyon scattering
amplitudes [13].  It turns out that these
can be found directly from the oscillator states
whose dynamics is exactly known.
Note first that the transition to
collective field theory seemingly introduces a degeneracy since one
could formally write separate states for the $\alpha_+$ and $\alpha_-$
sector.  The two sectors commute since the Poisson
brackets are $\{ \alpha_+ , \alpha_- \} = 0$.  For the
tachyon one would have two states

$$\int {dx\over 2\pi} \, {(\alpha_+ \pm x )\over 1\pm ik}^{1\pm ik}
\quad {\rm and} \quad \int {dx\over 2\pi} \, {(\alpha_- \pm x)\over
1\pm ik}^{1\pm ik}$$
with the same quantum numbers
$$po = k \,\, , \,\, p_x = -2 \pm ik$$
This would imply doubling which would be physically unacceptable.  The
resolution of the paradox is found in identifying these states with
each other [13].
This step can be (and is) thought of as a set of boundary conditions
imposed in the theory.

The above identification gives a nonlinear
relationship between left and right moving waves.  After a shift
by the classical background we have
$$\int_{\infty}^{\infty} d\tau \, e^{+ik\tau} {\hat{\alpha}_+\over
\mu}  = -
\int_{\infty}^{\infty} \, {d\tau\over ik\pm 1} \, e^{-ik\tau} \left[ \left(
1+ {\hat{\alpha}_-\over \mu}\right)^{ik+1} - 1 \right] $$
This is a solution to the scattering problem in the form first
found following a different route in [6]. Expansion of the above
equation generates general tree level N-point scattering
amplitudes. What we have seen is that the simple linear dynamics of
the matrix oscillator induces through collective phenomena the nonlinear
dynamics of string scattering.
\bigskip

\chapter{Symmetry}
\smallskip

The matrix model and the scalar field theory exhibit in an obvious
way a large symmetry
algebra.  Its fundamental origin lies in the symplectic structure
underlying the collective field theory where $M\rightarrow x$ and
$P\rightarrow \alpha (x,t)$ represent canonically conjugate variables.
The natural symmetry generators [8]
$$H_m^n = \Tr \left( P^m M^m \right)$$
generate general canonical transformations in the $M, P$ or
equivalently $x,\alpha$ phase space.  They close a $W_{\infty}$ algebra
$$\left[ H_{m_{1}}^{n_{1}} , H_M^{n_{2}} \right] = i \left( \left( m_2
- 1\right) n_1 - \left( m_1 - 1\right) n_2 \right) \, H_{m_{1} + m_{2}
- 2}^{n_{1} + n_{2}} $$

For the physical system given by the inverted oscillator one takes the spectrum
operators
$$ O_{JM} \equiv \Tr \left( P+M\right)^{J+M+1} \left(P-M\right)^{J-
M+1}$$
These in the field theory become
$$O_{JM} = \int \, {dx\over 2\pi} \int _{\alpha_{-}}^{\alpha_{+}}
d\alpha \left( \alpha + x \right)^{J+M+1} \left( \alpha - x\right)^{J-M+1}$$
and obey the $W_{\infty}$ commutation relations:
$$\left[ O_{J_{1}M_{1}} \, , \, O_{J_{2} M_{2}}\right] = - 4i
\left( \left( J_1 + 1\right) M_2 - \left( J_2 + 1\right) M_1 \right)
O_{J_{1} + J_{2} M_{1} + M_{2}}$$
The hamiltonian itself is a member of this algebra
$$H = O_{0,0} = \Tr \left( P^2 - M^2\right)$$
and this implies [8] that an infinite sequence of discrete states is
generated described by [9,10] two integers $J, M (-J \leq M \leq +
J)$.  In this sense the $W_{\infty}$ generators give a spectrum
generating algebra.  The nature of the imaginary energy discrete
states is nontrivial, in the present approach they are composite
states of the tachyon.

Analogous operators are found [11] in conformal field theory.  They are
given by the discrete state vertex operators
$$\Psi_{JM} = \left( H(z)\right)^{J-M} e^{(-1+J)\varphi(z)}$$
which with the corresponding ground ring operators
${\cal O}_{J,M}$ combine into a  conserved currents
$$ O_{JM} (z,\hat{z}) = \Psi_{J+1,M} (z) {\cal O}_{J,M} (\bar{z})$$

Returning to the collective approach one sees that the symmetry generators
$O_{J ???}$
directly represented as nonlinear functions of the tachyon field
$\alpha(x,t)$.  This immediately implies that the generators act in
a very special (nonlinear) way on the tachyon states.  Studying the
action of discrete vertex operators on the tachyon module, a similar
conclusion can be reached in the conformal calculus also.  This gives
rise to nonlinear Ward identities first described in [12].  These
identities are, however, explicitly encoded in the collective
representation.

We explore this in some detail [13].  First of all the exact tachyon
operators
$$T_n = \int \, {dx\over 2\pi} \, \int_{\alpha_{-}}^{\alpha_{+}}
d\alpha \left( \alpha \pm x \right)^n$$
can be written as extensions of the algebra $T_n = O_{{n\over2} -
1,n}$.  Of particular relevance is the subalgebra given by
$$O_{2N} \equiv O_{N,N} = \int \, {dx\over 2\pi} \, \int d\alpha \,
\left( \alpha + x\right)^{2N+1} \left( \alpha - x\right)$$
Since it represents a Virasoro algebra:
$$ \left[ O_{2N} , O_{2N'}\right] = 4i \left(N-N'\right)
O_{2(N+N')}$$
The Virasoro generators act on the tachyon operators $T_n$ in the
simple way
$$\left[ O_{2N} , T_n \right] = 2in \, T_{n+2N}$$
These commutators can be directly verified from the collective representation.
Consequently the tachyon field can be thought to have (space--time)
conformal spin 1 (in the next section this notion and its
generalization will be pursued further).

To exhibit the higher Ward identities, one goes to
the approximation with the vanishing cosmological
constant. In conformal field theory it is actually only this limit that is
well understood.  The background shift for the
collective fields now becomes
$$\alpha_{\pm} = \pm \sqrt{\tilde{x} - \mu} + \bar{\alpha}_{\pm}
\rightarrow \alpha_{\pm} = \pm x + \bar{\alpha}_{\pm}$$
and after a change $x = e^{-\tau}$ the generators become
$$\eqalign{ O_{JM} &  \approx {2^J\over J-M+2} \int {dt\over 2\pi}
e^{2Mi\tau} \alpha_+^{J-M+2}  \cr
& + {2^J\over J+M+2} \int {dt\over 2\pi} \, e^{-2Mi\tau} \, \alpha_-
^{J+M+2} }$$

Expanding in terms of linearized tachyon creation--annihilation
operators $\bar{\alpha}_+ (k) = \alpha(k), \alpha_- (k) = \beta(k)$
we have the expression
$$\eqalign  {O_{JM} & = {2^J\over J-M+2} \, \int dk_1 dk_2 \cdots dk_{J-
M+2} \alpha (k_1) \cdot \alpha (k_{J-M+2}) \delta \left( \sum k_i +
2M\right)\cr
& + {2^J \over J+M+2} \int  dp_1 \, dp_2 \cdots dp_{J+M+2} \beta (p_1 )\cdots
\beta \left( p_{J+M+2}\right) \delta \left( \sum p_i + 2M\right) }$$
These are similar (but not identical) to the representation of the
symmetry generators constructed on the linearized
tachyon by Klebanov in [12].
Some subtle differences are explained in [13].  The implications when
actioning on tachyon states are all the same, one has
$$O_{M+N,M} \vert k_1 k_2 \cdots k_{N+1} \rangle = \left( M + \sum_i
k_i\right) \vert k\sum_i k_i + M\rangle$$
This nontrivial identity is sufficient to construct the bulk
scattering applitudes as was done in [12].

We have explained how the scalar collective field representation
induces the same Ward identities as the conformal vertex operator
calculus.  One can in fact invert this and realize that the collective
formulation gives the simplest realization of the Ward identities.
An advantage of the field theoretic approach is the fact that the nonzero
cosmological constant is easily incorporated.  It represents a
nonzero chemical potential in the field theory.  An exact discussion of Ward
identities can now also be contemplated.  The symmetry generators can be
completely defined through normal ordering over the field oscillators
as has been done in detail in [3] for the hamiltonian.

To summarize the above discussion we have seen that the $W_{\infty}$
symmetry occurs naturally in the canonical framework of the matrix
model and collective field theory.  These are based on phase space
notions $(M, P)$ and $(x, \alpha (x,t ))$ respectively and the
symmetry is given by general canonical transformations.  It is an
interesting question to ask what happens to these at the full quantum
level.  In general canonical transformations are known to be
highly nontrivial in quantum mechanics.
\bigskip
\chapter{Extension:  Coupling of Topological and Collective Fields}
\smallskip

We have seen that the $W_{\infty}$ symmetry seems
to govern the dynamics of the theory.  Through Ward
identities and its representation in terms of the tachyon field one has a
nonlinear realization giving the complete c=1 tachyon S--matrix.  This
is manifest in the collective description where the $W_{\infty}$
generators are explicit nonlinear functions of the scalar field.  The
generators
play a role of a spectrum generating algebra giving a sequence of
additional discrete states.  These appear as composite
states [8] of the collective hamiltonian.  One
would then like to introduce separate fields [11] to represent these states.
This would allow a study of nontrivial backgrounds where these fields
play a role.  One such example would be the black hole.

The question of how
the discrete higher string fields should be introduced is a nontrivial
one.  On one hand one has the standard BRST gauge symmetry while in the
matrix model one finds a $W_{\infty}$ symmetry.
It is suggestive to follow the latter as a fundamental
principle, and develop an interacting theory of collective and higher
fields in two
dimension based on $W_{\infty}$.  The physical nature of the two
sets of fields is, however, quite
distinct:  the tachyon is a full dynamical particle with the
associated scalar field while the higher modes are global in structure
and should be of topological nature.  It is useful then to introduce
[16] the notion of space--time (as opposed to world sheet) central charges.
With this we find that the collective tachyon field can be characterized as
carrying the central change c=1 while the infinite sequence of higher
fields all carry c=0 and are therefore topological.  To define the
coupling we begin [16] with a quantum c=1 $W_{1+\infty}$ algebra defined
for example through conformal fermi fields:
$$B^k (z) = : \psi^+ (z) \partial_z^k \psi (z):$$
The algebra of commutators (with central charges corresponding to c=1)
reads
$$\eqalign{ \left[ B_n^0 , B_m^0\right] & = n \delta n+m_0\cr
\left[ B_n^1 , B_m^0\right] & = - m B_{m-n}^0 + {n(n-1)\over 2}
\delta_{n+m}\cr
\left[ B_n^1 , B_n^1\right] & = (n-m) B_{n+m}^1 - {n^3 -n\over 6}
\delta_{n+m}\cr
\left[ B_n^2 , B_m^0 \right] & = m^2 B_{n+m}^0 - 2m B_{n+m}^1 +
{n(n-1)(2n-1)\over 6}\cr
& \vdots }$$

The idea is then to realize the algebra in terms of bosonic fields.
In the first commutator we recognize the scalar collective field
$$ \alpha (z) \equiv B^0 (z) \quad\quad \left\{ \alpha (x), \alpha
(x') \right\} = \partial_x \delta (x-x')$$

One then ``solves" the commutators for the next $w_{\infty}$ generator
$B^1$ finding
$$B_n^1 = : {1\over 2} \alpha_{n-\ell} \alpha_{\ell} : + {n-1\over 2}
\alpha_n + w_n^1$$
Here the scalar field is seen to be
responsible for the total central change
and a remnant spin 2 field appears with
$$\left[ w_n^1 , w_m^1\right] = (n-m) w_{n+m}^1$$
It consequently has c=0 and commutes with the collective boson
$[\alpha_n, w_m^1] =0$.  The strategy is then clear, we continue with
the higher commutators and find a sequence of topological objects.  In
particular for the next generator the commutators are uniquely solved by
$$B_n^2 = {1\over 3} : \alpha^3 : + 2 \sum w_{n-\ell}^1 \alpha_{\ell}
+ w_n^2$$
where $w_n^2$ represents a spin 3 topological field.
This generator gives the hamiltonian which
now reads
$$H = \int dx \left\{ {1\over 12\pi} \alpha^3 (x,t) + w_1 (x,t) \alpha
(x,t) + w_2 (x,t)\right\}$$

One has an interaction with the collective scalar
with a unique coupling given by $\alpha (x,t ) w_1 (x,t)$.  This is an
interaction of the spin 1 (collective) and spin 2 (Virasoro) fields.

  The infinite component
$W_{\infty}$ symmetry is likely to play a role
The higher spin fields $\left\{w_n (x,t) ,
n=1,2,\cdots\right\}$ are found to obey the centerless $w_{\infty}$
algebra

$$\left[ w_n (x) , w_m (y) \right] = \left( (n-1) w_{n+m} + (M-1)
w_{n+m} (y) \right) \partial_x \delta (x-y)$$
These fields are then topological.  There is actually a representation
which makes this manifest.  Consider an infinite component
field defined through a
general pseudo differential operator
$$K (x,\partial) = 1 + \sum_{j=1}^{\infty} a_j (x,t) \partial^{-j}$$
This gives a representation where the hamiltonian densities are all total
divergences [17]:
$$W_n (x,t) = Res \left( K^{-1} \partial^n K\right) = \partial_x
\left( \sigma_n \right)$$
This form also exhibits the topological modes as pure
gauges of a large $W_{\infty}$ gauge group.  $K$ can be
thought of as a group element and $ A=K^{-1} \partial K$ as a gauge
field.  It is
likely that a gauge invariant $ W_{\infty}$ Lagrangian can be written which
after gauge fixing reduces to the above.

One expects that the relevance of higher fields will come for studying
other classical backgrounds.  The formalism discussed in much of this
talk is clearly defined around the flat dilaton vacuum.  Based on
conformal field theory insights, one expects a black hole to
arise as a nontrivial background
(see recent suggestions [19-21]).

\bigskip
\noindent{\bf Acknowledgement:}

It is a pleasure to thank the organizers of the Nishinomiya-Yukawa
meeting, in particular Dr. M. Ninomiya for a most memorable meeting.
\endpage

\centerline{\it REFERENCES}
\bigskip

\pointbegin
D. J. Gross and A. A. Migdal, {\it Phys. Rev. Lett.} {\bf 64} (1990)
127;
M. R. Douglas and S. Shenker, {\it Nucl. Phys.} {\bf B335} (1990) 635;
E. Br\'ezin and V. Kazakov, {\it Phys. Lett.} {\bf B236} (1990) 144;
D. J. Gross and N. Miljkovi\'c, {\it Phys. Lett.} {\bf B238} (1990) 217;
E. Br\'ezin, V. A. Kazakov and A. B. Zamolodchikov, {\it Nucl.Phys.}
{\bf B338} (1990) 673;
P. Ginsparg and J. Zinn-Justin, {\it Phys. Lett.} {\bf B240} (1990)
333.
\point
S. R. Das and A. Jevicki, {\it Mod. Phys. Lett.} {\bf A5} (1990) 1639;
A. Jevicki and B. Sakita, {\it Nucl.Phys.} {\bf B165} (1980) 511.

\point
K. Demeterfi, A. Jevicki and J. P. Rodrigues, {\it Nucl.Phys.}
{\bf B362} (1991) 173; {\bf B365} (1991) 499; {\it Mod. Phys. Lett.}
{\bf A35} (1991) 3199.
\point
J. Polchinski, {\it Nucl.Phys.} {\bf B362} (1991) 25.
\point
D. Gross and I. Klebanov, {\it Nucl. Phys.} {\bf B352} (1991) 671;
A. M. Sengupta and S. Wadia, {\it Int. J. Mod. Phys.} {\bf A6} (1991)
1961;  G. Moore, {\it Nucl. Phys.} {\bf B368} (1992) 557.
\point
G. Moore and R. Plesser, \lq\lq Classical Scattering in 1+1 Dimensional
String Theory", Yale preprint YCTP-P7-92, March 1992.
\point
P. D. Francesco and D. Kutasov, {\it Nucl. Phys. B} (1991).
\point
J. Avan and A. Jevicki, {\it Phys. Lett.} {\bf B266} (1991) 35;
{\bf B272} (1992) 17;
D. Minic, J. Polchinski and Z. Yang, {\it Nucl. Phys.} {\bf B369}
(1992) 324;
G. Moore, N. Seiberg; {\it Int. J. Mod. Phys.} {\bf A7} (1992) 2601;
S. R. Das, A. Dhar, G. Mandal and S. Wadia;
   {\it Mod. Phys. Lett.} {\bf A7} (1992) 71; V. H. Danielsson, {\it
Nucl. Phys.} {\b380} (1992) 183.
\point
D. Gross, I. Klebanov and M. Newman, {\it Nucl. Phys.} {\bf B350}
(1991) 621; A. M. Polyakov, {\it Mod. Phys. Lett.} {\bf A6} (1991) 635;
Lectures given at 1991 Jerusalem Winter School.
\point
B. Lian and G. T. Zuckermon, {\it Phys. Lett.} {\bf B254} (1991); K.
Itoh and N. Ohta, Fermilab preprint (1991); P. Bouwknegt, T. McCarthy
and K. Pilch, CERN preprint (1991).
\point
E. Witten, {\it Nucl. Phys.} {\bf B373} (1992) 187;
I. Klebanov and A. M. Polyakov, {\it Mod. Phys. Lett} {\bf A6}
(1991) 3273;
N. Sakai and Y. Tanii, {\it Prog. Theor. Phys.} {\bf 86} (1991) 547;
J. Barbon, {\it Int. Journal of Mod. Phys.} {\bf Vol.7} (1992) 7579.
\point
I. R. Klebanov, \lq\lq Ward Identities in Two-Dimensional String
Theory", {\it Mod. Phys. Lett.} {\bf A7}, (1992) 55.
D. Kutasov, E. Martinec and N. Seiberg, PUPT-1293, RU-31-43;
E. Witten and B. Zweibach, {\it Nucl. Phys.} {\bf B377} (1992) 55.
\point
A. Jevicki, J. P. Rodrigues and A. J. van Tonder, ``Scattering States
and symmetries in the Matrix Model and Two Dimensional String Theory",
Brown-Het-874, (1992).
\point
A. Dhar, G. Mandal and S. Wadia, {\it Mod. Phys. Lett.} {\bf A}
(1992).
\point
S. Iso, D. Karabali and B. Sakita, {\it Nucl. Phys.} {\bf B} (1992).
\point
J. Avan and A. Jevicki, ``Interacting Theory of Collective and
Topological Fields in 2 Dimensions", (to appear in Nucl. Phys. B).
\point
A. Jevicki and T. Yoneya, {\it Mod. Phys. Lett.} {\bf A5} (1990) 1615.
\point
M. Li, ``A Proposal onthe Topological Sector of 2D String",
Brown preprint (December 1992).
\point
J. G. Russo, UTTG-27-92 November 92; \nextline
S. R. Das, TIFR-TH92-162; \nextline
A. Dhar, G. Mandal and S. Wadia, TIFR-TH92-163; \nextline
T. Yoneya, UT-Komaba-92-
113.
\point
E. Martinec and S. Shatashvili, {\it Nucl. Phys.} {\bf B368} (1992)
338.
\point
S. Mukhi and C. Vafa, Harvard/Tata preprint (January 1993).
\endpage

\end